\documentclass{article}
\usepackage{graphics}
\title{\bf Can black holes exist in an accelerating universe?}
\author{ Pedro F. Gonz\'{a}lez-D\'{\i}az.\\
Instituto de Matem\'{a}ticas y F\'{\i}sica Fundamental\\ Consejo Superior
de Investigaciones Cient\'{\i}ficas\\ Serrano 121, 28006 Madrid,
SPAIN\\ }
\date{May 15, 2002}
\begin{document}
\maketitle \large \setlength{\baselineskip}{0.9cm}
\begin{center}
{\bf Abstract}
\end{center}
This letter deals with an analysis of the space-time static metric
that corresponds to a quintessential state equation with constant
characteristic parameter. Following a procedure parallel to as it
is used in the case of de Sitter space, we have tried to
generalize the metric components that correspond to the
quintessential case to also embrace black-hole terms and shown
that this is not possible. We argue therefore that, in the absence
of a cosmological constant, black holes seem to be prevented in a
cosmological space-time which is asymptotically accelerating when
the acceleration is driven by a quintessence-like field.

\vspace{5cm}

\noindent Report-no: IMAFF-RCA-02-05

\pagebreak

Some time ago it was not infrequent to witness scientific debates,
both at meetings and private desks, on whether black holes may
actually exist. Many physicists were then reluctant to accept
black holes to indeed be among the existing physical objects. Even
Einstein showed a very critical attitude against the existence of
black holes and advanced some proofs for their nonexistence [1].
Since then, the situation has dramatically shifted to favor the
opinion that black holes must exist and be formed by gravitational
collapse in many places of the universe, including star binaries
and the center of galaxies. Although nobody has actually found a
completely convincing case in which the black-hole occurrence is
out of any doubt [2], black holes are currently being also used as
the engine to deliver the high energies observed in many
astrophysical processes, and often claims that a black hole has
been observed in binaries or at galactic centers are being
released by the press.

Basing on the gravitational physics induced by a cosmic
quintessential field [3], we argue in this letter that the present
accelerating expansion of the universe [4,5] may not be compatible
with the occurrence of large black holes, provided that the
accelerating expansion is driven by a dynamic vacuum scalar field
with negative pressure which did not imply a future cosmic event
horizon [6,7] and, therefore, that the fundamental string and M
theories be mathematically consistent [8]. Even in this case, room
is left however for the existence of primordial black holes [9]
during the whole cosmic period from big bang to the onset of the
accelerating regime. If the recent results in supernova type Ia
[4,5] are nevertheless explained by means of a positive
cosmological constant, any kinds of black holes are allowed, but
then no string or M theory could be consistently defined. Our
result becomes then a kind of cosmic dilemma by which one must
somehow choose between black holes and fundamental string or M
theories.

The Schwarzschild solution is usually interpreted as describing a
black hole of given mass $M$ in an asymptotically flat space.
There is also the straightforward generalization to the case of
nonzero positive cosmological constant $\Lambda$, which is
provided by the so-called Schwarzschild-de Sitter static metric
[10,11], which represents a black hole in asymptotically de Sitter
space [12]. This solution reads
\begin{equation}
ds^2= -\left(1-\frac{2GM}{r}-\frac{\Lambda r^2}{3}\right)dt^2
+\left(1-\frac{2GM}{r}-\frac{\Lambda r^2}{3}\right)^{-1}dr^2+r^2
d\Omega_2^2 ,
\end{equation}
where $d\Omega_2^2$ is the metric on the unit three-sphere. This
generalization of Schwarzschild space can actually be also viewed
as a generalization of de Sitter space in static coordinates
because of the following argument. If we assume for the considered
static metric a general spherically-symmetric {\it ansatz}
\begin{equation}
ds^2= -A(r)dt^2 +B(r)dr^2+r^2 d\Omega_2^2 ,
\end{equation}
in the case of a de Sitter space the metric components
$A(r)=B(r)^{-1}$ satisfies a differential equation which is the
particular case for $\omega=-1$ of the most general differential
equation that corresponds to a static, spherically-symmetric space
for a quintessence-like vacuum scalar field with equation of state
$p=\omega\rho$ [3,13]
\begin{equation}
\left(\frac{A'
A^{-(1+\omega)/(2\omega)}}{r^{(1+2\omega)/\omega}}\right)'
+\left(\frac{A^{(\omega-1)/(2\omega)}}{r^{(1+
3\omega)/\omega}}\right)' +\frac{\alpha\left(A' r^2\right)'}{8\pi
G\omega r^{(1+6\omega)/\omega}}=0 ,
\end{equation}
where the prime $'$ denotes differentiation with respect to the
radial coordinate $r$, $\alpha$ is an integration constant and the
state equation parameter $\omega$ runs in the interval
$-1/3\geq\omega\geq -1$, and either [13]
\begin{equation}
B=1
+\frac{\left[(A')^{-2\omega/(1+\omega)}Ar^2\right]'}{A(A')^{-(3\omega
+1)/(\omega+1)}\left(A' r+\frac{2(3\omega+1)A}{\omega+1}\right)} ,
\end{equation}
for $-1/3 >\omega > -1$, or $B=A^{-1}$ for $\omega=-1$. Note that
Eqs (3) and (4) admit a particular solution for any $\omega$ other
than $\omega=-1$ which has the form $A\propto r^{-1}$,
$B=1-r^{-2}$, and is singular at $r=0$ but does not show any event
horizon.

At the extreme de Sitter case where $\omega=-1$ for which the
positive cosmological constant is given by $\Lambda=8\pi
G/\alpha$, Eq. (3) reduces to
\begin{equation}
\left(\frac{A'}{r}\right)' +\left(\frac{A}{r^2}\right)'
=\frac{1}{\Lambda r^5}\left(A' r^2\right)' .
\end{equation}
Of course, a solution to Eq. (4) is the well-known static de
Sitter metric $A=B^{-1}=1-\Lambda r^2/3$. In addition, there is
the generalization of this solution to a de Sitter space-time
containing a black hole when we make the transformation
$A\rightarrow\tilde{A}=A(r)+\xi(r)$, where $\xi(r)$ is also a
function of the radial coordinate $r$ that satisfies the following
two conditions separately
\begin{equation}
\left(\frac{\xi '}{r}\right)' +\left(\frac{\xi}{r^2}\right)'
=0,\;\; \xi ' r^2={\rm const.} .
\end{equation}
Eq. (5) is then invariant under $A\rightarrow\tilde{A}$ if
$\xi={\rm const.}/r$, so that we finally obtain the
Schwarzschild-de Sitter solution
$\tilde{A}=\tilde{B}^{-1}=1-\Lambda r^2-2GM/r$, after fixing ${\rm
const.}=-2GM$ in the limit $\alpha\rightarrow\infty$.

In order to investigate whether there could exist black holes in
asymptotically accelerating cosmological spaces corresponding to a
dynamic vacuum with a quintessence-like equation of state
characterized by a constant parameter $-1/3 >\omega > -1$, rather
than $\omega=-1$ of de Sitter space, let us try to follow a
procedure parallel to that we have used for the case that we have
just a positive cosmological constant, starting also with Eq. (3).
We first notice that in all cases where $\omega\neq -1,-1/3$, this
general differential equation can be cast in a simpler form if we
(i) re-define $A(r)$ such that
\begin{equation}
D(r)=A(r)^{(\omega-1)/(2\omega)} ,
\end{equation}
and (ii) use the general solution to Eq. (3) for $-1/3 >\omega
>-1$, i.e. [13]
\begin{equation}
D(r)= \kappa r^{2(\omega-1)/(\omega+1)} ,
\end{equation}
where
\begin{equation}
\kappa\equiv\kappa(\omega,\alpha) = \left(\frac{2\pi
G\left(\omega^2+6\omega+1\right)}{\omega\alpha}\right)^{(\omega-1)/(\omega+1)}
.
\end{equation}
In terms of the new function $D(r)$, the differential equation (3)
can then be re-written as
\begin{equation}
\left(\frac{D'}{r^{(2\omega+1)/\omega}}\right)'
+\left(\frac{\omega-1}{2\omega}\right)\left(\frac{D}{r^{(3\omega+
1)/\omega}}\right)'+\frac{\alpha\kappa^{(\omega+1)/(\omega-1)}}{8\pi
G\omega r^{(6\omega+1)/\omega}}\left(D' r^4\right)' = 0.
\end{equation}
This equation is obviously satisfied by solution (8). What we
shall try to see now is whether it will be also satisfied by the
generalized function $\tilde{D}$, defined by
\begin{equation}
D\rightarrow\tilde{D}=D(r)+\psi(r) ,
\end{equation}
in such a way that, analogous to the function $\xi(r)$ in de
Sitter space, the function $\psi(r)$ will satisfy the following
two conditions:
\begin{equation}
\left(\frac{\psi '}{r^{(2\omega+1)/\omega}}\right)'
+\left(\frac{\omega-
1}{2\omega}\right)\left(\frac{\psi}{r^{(3\omega+1)/\omega}}\right)'
=0 , \;\; \psi ' r^4={\rm const.} ,
\end{equation}
so that Eq. (11) is satisfied by the solution $\psi(r)$ too. It
can however be straightforwardly checked that for the two
conditions (12) to be simultaneously satisfied by $\psi(r)$ it is
necessary that $(\omega-1)/(2\omega) = 3$, which can only hold for
$\omega=-1/5$, i.e. outside the range that corresponds to the
accelerating solutions $-1/3 >\omega
> -1$. This shows that, even though for $\omega=-1$ the
de Sitter (or Schwarzschild) static solution can be consistently
generalized to the Schwarzschild-de Sitter metric, so allowing
black holes to exist in asymptotically de Sitter space,
Schwarzschild black holes are {\it not} allowed to occur in
universes which asymptotically tend to a spherically symmetric
space that corresponds to the accelerating cosmological spaces
whose expansion is driven by a quintessence-like field with an
equation of state characterized by a constant parameter $-1/3
>\omega >-1$. It is worth noticing that even for the case
$\omega=-1/5$ we had no Schwarzschild event horizon at all, as for
in that case we obtain $\psi\propto D\propto r^{-3}$.

The question now arises: if the present accelerated expansion of
the universe is driven by a quintessence field, is then the above
result a definitive proof for the nonexistence of Schwarzschild
black holes (or actually any of their generalizations with charge
or angular momentum) in the accelerating universe we live in?.
Even though the considered values of $\omega$, $-1/3
>\omega
>-1$, strictly make the usual sense only for homogeneous and
isotropic Friedmann-Robertson-Walker (FRW) spacetimes, if we keep
an equation of state $p=\omega\rho$ also in the case of spacetimes
with static, spherically symmetric coordinates, one can as well
obtain the static metrics which correspond to the above
$\omega$-values [13] for a given fixed relation between the energy
density and the metric components, much in the same way as the
static metric for de Sitter or Schwarzschild-de Sitter spaces can
be derived from the Einstein equations for static, spherically
coordinates and an equation of state $p=\omega\rho$, whenever we
set $\omega=-1$. That static metric for de Sitter space can, in
fact, be directly related with the corresponding FRW de Sitter
metric obtained from the Friedmann equations also for $\omega=-1$
by means of an embedding in a common five-dimensional hyperboloid.
Thus, to the extent at which Schwarzschild-de Sitter metric be
interpreted as representing a black hole in asymptotically de
Sitter space, our result should in fact imply a positive answer to
the above question. Schwarzschild, Kerr and Reissner-Nordstr\"{o}m
black holes could not then exist in an accelerating universe.
Moreover, since black holes with sizes larger than that of the
atomic scale would completely evaporate off by the Hawking process
only in a time which is much longer than the present age of the
universe, it appears that for quintessence-driven accelerated
expansion none of the known black holes with masses larger than
$\sim 10^{15}$ g could exist at any time along the entire
cosmological evolution, from big bang to now.

However, even though neither Schwarzschild event horizon nor any
of its rotating or charged generalizations can exist in an
accelerating universe, what such a universe might still have is a
new kind of generalized horizons which would in turn be forbidden
for the case $\omega=-1$. These possibilities would correspond to
those remaining generalizations which, together with
$A\rightarrow\tilde{A}$, $\tilde{B}=\tilde{A}^{-1}$, will exhaust
all possible generalizations for the type of spacetimes we are
considering. Let us introduce then the additional transformation
\begin{equation}
D\rightarrow\bar{D}=D(r)\exp[\chi(r)] .
\end{equation}
Insertion of this transformation into the differential equation
(10) and further separate application of this equation to $D(r)$
alone leads then to:
\begin{equation}
\exp[\chi(r)]= C_0
+C_1\left(\frac{\omega+1}{2(\omega-1)}\right)\int
x^{\frac{m(\omega+1)}{2(\omega-1)}-\left(\frac{\omega-3}{\omega+1}\right)}
e^{-x}dx ,
\end{equation}
where $C_0$ and $C_1$ are arbitrary integration constants, and we
have introduced the definitions
\[x=r^{2(\omega-1)/(\omega+1)} ,\;\;\;
m=\frac{10\omega^2+30\omega+4}{3\omega^2-6\omega+1} .\] It can be
checked that (i) for $\omega=-1$ Eq. (14) would imply
$\chi=$const., meaning that no solution other than de Sitter or
Schwarzschild-de Sitter are allowed for $\omega=-1$, and (ii) for
$-1/3 >\omega >-1$ the function $\chi$ does actually depend on $r$
and therefore may, in principle, eventually show new kinds of
event horizons. Eq. (14) admits an integration in terms of the
incomplete gamma function ${\bf \gamma}(\alpha,x)$ [14], such that
\begin{eqnarray}
&&\exp(\chi)=C_0+C_1\left(\frac{\omega+1}{2(\omega-1)}\right) {\bf
\gamma}\left[\frac{m(\omega+1)}{2(\omega-1)},x\right]\nonumber\\
&&=C_0+C_1\left(\frac{\omega+1}{2(\omega-1)}\right) {\bf
\gamma}\left[\frac{m(\omega+1)}{2(\omega-
1)},r^{\frac{2(\omega-1)}{\omega+1}}\right] .
\end{eqnarray}
Hence, we can obtain an expression for the metric component $A(r)$
which reads
\begin{eqnarray}
&&A(r)=\left[\bar{D}(r)\right]^{\frac{2\omega}{\omega-1}}
=\nonumber\\ &&\kappa^{\frac{2\omega}{\omega-1}}
r^{\frac{4\omega}{\omega+1}}\left\{C_0
+C_1\left(\frac{\omega+1}{2(\omega-1)}\right){\bf
\gamma}\left[\frac{m(\omega+1)}{2(\omega-1)},
r^{\frac{2(\omega-1)}{\omega+1}}\right]\right\}^{\frac{2\omega}{\omega-1}}
.
\end{eqnarray}
Inspection of Eqs. (2), (4) and (8) allows us to deduce that all
possible singularities of the metric must arise from the following
two conditions: (i) $A=0$, and (ii)
$A'r+2(3\omega+1)A/(\omega+1)=0$, but not from $A'=0$ alone, in
such a way that either $A=0$ and $B=\infty$ simultaneously (i.e. a
coordinate singularity at an event horizon) or $A\rightarrow
B\rightarrow\infty$ (i.e. a curvature singularity which was
already present before making generalization (13)). Let us first
analyze condition (ii), noting that the $r$-derivative of $A(r)$
can be written as [15]
\begin{equation}
A(r)' =
\kappa^{\frac{2\omega}{\omega-1}}r^{\frac{3\omega-1}{\omega+1}}
\left\{\frac{4\omega}{\omega+1}-C_1\exp\left[-\left(\chi+
r^{\frac{2(\omega-1)}{\omega+1}}\right)\right]\right\}
\exp\left(\frac{2\omega\chi}{\omega-1}\right) .
\end{equation}
Using then expression (17) in condition (ii) we get for the
possible singularities at $r=r_h$ and $r=r_s$
\begin{eqnarray}
&&\{P(r_h)\}\times\{Q(r_s)\} \equiv
\left\{r_h^{\frac{4\omega}{\omega+1}}\exp\left(\frac{2\omega\chi(r_h)}{\omega-
1}\right)\right\}\times \nonumber\\
&&\left\{\frac{2(5\omega+1)}{\omega+1}
-C_1\exp\left[-\left(\chi(r_s)+r_s^{\frac{2(\omega-1)}{\omega+
1}}\right)\right]\right\}= 0.
\end{eqnarray}
We have then two different possible singularities coming from
condition (ii). If we set $\{P(r_h)\}=0$, then the use of
definitions (15) leads to $r_h=\infty$, so that, since $r\geq 0$
and in the considered $\omega$-interval $2(\omega-1)/(\omega+1)$
is definite negative and $m(\omega+1)/[2(\omega-1)]$ is definite
positive, we also have ${\bf \gamma}(r_h)=0$ and $A(r_h)=0$, which
makes conditions (i) and (ii) to hold simultaneously. Such a
singularity would then describe a black hole or cosmological event
horizon at $r_h=\infty$, and hence the resulting static spacetime
would correspond to an accelerating universe either within an
asymptotic black hole space with infinite radius, or with a
cosmological event horizon at $r=\infty$. It follows that the
curvature singularity at $r_s=0$, at which both metric components
$A$ and $B$ simultaneously tend to infinity, must correspond to
the condition $\{Q(r_s)\}=0$. This would in turn imply
\begin{equation}
\frac{2(5\omega+1)}{\omega+1}C_0
=C_1\left[(\omega+1)\exp\left(-r_s^{\frac{2(\omega-1}{\omega+1}}\right)-
\left(\frac{5\omega+1}{\omega-1}\right){\bf \gamma}(r_s)\right] ,
\end{equation}
where [15]
\begin{equation}
\lim_{r\rightarrow 0}{\bf \gamma}(r)
=\Gamma\left[\frac{m(\omega+1)}{2(\omega-1)}\right],
\end{equation}
with $\Gamma$ being the complete gamma function. One can now set
the relative value of the integration constants $C_0$ and $C_1$ to
be
\begin{equation}
C_0=\frac{(\omega+1)\Gamma\left[\frac{m(\omega+1)}{2(\omega-
1)}\right]C_1}{2(1-\omega)} ,
\end{equation}
with which we finally obtain for the metric components
\begin{equation}
A(r)=\kappa^{\frac{2\omega}{\omega-1}}
r^{\frac{4\omega}{\omega+1}}\left[\frac{(\omega+1)C_1}{2(1-
\omega)}\right]^{\frac{2\omega}{\omega-
1}}\left\{\Gamma\left[\frac{m(\omega+1)}{2(\omega- 1)}\right]
-{\bf \gamma}\left[\frac{m(\omega+1)}{2(\omega-1)},
r^{\frac{2(\omega-1)}{\omega+1}}\right]\right\}^{\frac{2\omega}{\omega-1}}
.
\end{equation}
\begin{equation}
B(r)=\frac{F(r,\omega)}{A(r)} ,
\end{equation}
where $F(r,\omega)$ is a generally finite function of $r$ and
$\omega$ which diverges only as $r\rightarrow 0$ and as
$r\rightarrow\infty$.

We have in this way obtained that the generalization (13) implies
the existence of a spacetime which can be interpreted as
representing a space that corresponds to an accelerating universe
which evolves either within an infinite black hole, or toward an
asymptotic cosmological event horizon at infinity. In both of
these interpretations fundamental string and M theories would
preserve mathematical consistency as such interpretations do not
allow for causally disconnected regions anywhere up to infinite
future. This result encompasses all $\omega$-values within the
interval $-1/3 >\omega >-1$ and confirms therefore our above
conclusion that neither cosmological nor black hole event horizons
can exist at finite distances for state equations $p=\omega\rho$
with $-1/3 >\omega >-1$.

Thus, the only black holes which are allowed to occur within the
universe before the onset of the accelerating regime in the above
scenario are those having a mass scale $M\leq 10^{15}$ g. These
tiny black holes are currently known as primordial black holes [9]
and were formed quite before galaxies existed. They could have
evaporated off completely before or just at the coincidence time
[16], when dark energy started to dominate (see Fig. 1). In fact,
the black hole evaporation loss rate is given by the approximate
expression [17]
\[\frac{dM}{dt}\simeq - T_{BH}^4\times A_{BH}\sim \frac{\nu}{M^2} ,\]
where $T_{BH}=1/(8\pi GM)$ is the Hawking temperature of the black
hole and $A_{BH}=16\pi G^2 M^2$ is the black hole surface area.
The constant $\nu$ is expected to be a number of order unity which
reflects the number of degrees of freedom accounting for the
species of particles created by the black hole gravitational
field. Thus, an estimate of the time lasted by the largest
possible black hole in completely evaporating, $t\sim 64 G^2 M^3$,
should then roughly equal the coincidence time. This would bound
the mass of the largest possible black hole to be of the order
$10^{15}$ g and moreover imply that short gamma ray bursts which
were assumed by Cline, Motthey and Otwinowski [18] to be produced
by nearby primordial black holes had necessarily stopped to occur
at the onset of the accelerating regime. It would then appear that
the onset of dark energy dominance and the end of black hole
occurrence are strongly correlated to each other. At present, we
had no explanation for that correlation.

\vspace{1cm}

\begin{figure}[h]
\begin{center}
\scalebox{.6}[.5]{\includegraphics*{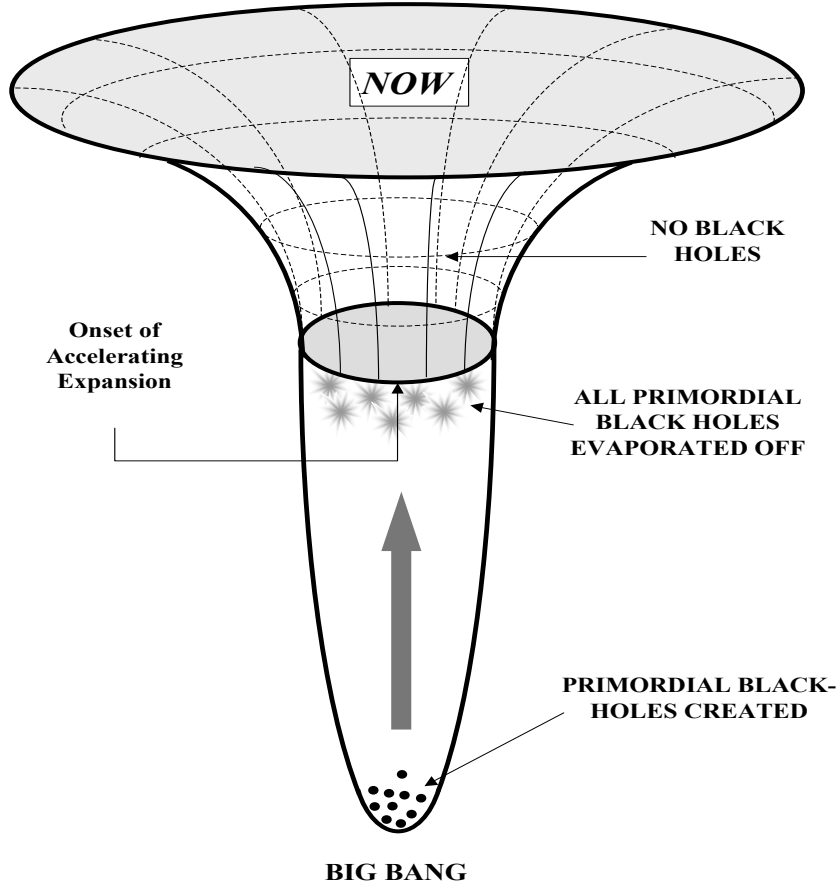}}
\end{center}
\caption{Pictorial representation of the evolution from big bang
to now of a universe which contains a quintessence field with
constant state equation responsable for the present accelerating
expansion. No black hole at all may exist in such a universe after
the onset of the present accelerating expansion. Before the
coincidence time at which dark energy started dominating, that
universe only allows the existence of primordial black holes which
exhausted all their energy by Hawking evaporation before or at the
onset of the present accelerating expansion.}
\end{figure}

If we want to preserve consistency in the fundamental theories for
gravity and particle physics, such as predictive string theories
or M theory, one cannot describe our present accelerating universe
using a gravitational framework with a positive cosmological
constant, for in that case the above theories become
mathematically ambiguous because they all rely on the existence of
an S matrix at infinite distances which is incompatible with the
existence of a cosmological event horizon in the future [6-8]. The
accelerating universe that corresponds to a constant
quintessential field has nevertheless no event horizons at finite
distances [13] and therefore does not pose this kind of challenge
to the fundamental theories. If we accept the result obtained in
this paper, then one is thus brought to the following dilemma:
either one allows for the existence of black holes of any size at
any time during cosmological evolution at the cost of having to
renounce to any mathematically consistent description of particle
physics and quantum gravity based on the existence of an S matrix
at infinite distances [6-8], or, if the mathematical consistency
of such teories is taken to be prioritary, then one has to
renounce to having black holes at our disposal to make the nuclei
of galaxies, or for igniting the engines of so many high-energy
processes that are currently thought to occur in astrophysics and
cosmology.

Although the present status of string and M theories, on the one
hand, and that of accelerating cosmology, on the other hand, does
not allow one to succeed in finding any solid, definitive argument
in favor of any of these two possibilities, one might be more akin
to preserving the predictive character and mathematical
consistency of fundamental particle-physics and quantum-gravity
(if any) theories, and then reluctantly shift the black-hole
concept to just the quantum realm where it could be implemented
either (i) as the kind of primordial black holes we considered
before, or, together with other space-time constructs such as
wormholes, ringholes or warp drives, as just another more
component of the so-called quantum space-time foam [19]. What this
attitude would definitely prevent is any large black hole formed
by the gravitational collapse of stars.

\vspace{.8cm}

\noindent{\bf Acknowledgements} The author thanks C.L. Sig\"uenza
for many useful discussions. This work was supported by DGICYT
under Research Project No. PB98-0520.

\pagebreak

\noindent\section*{References}

\begin{description}
\item [1] A. Einstein, Ann, Math. 40 (1939) 992.
\item [2] R. Narayan and J.S. Heyl, {\it Evidence for the Black
Hole Event Horizon}, to appear in Proceedings of the {\it Coral
Gable Conference on High Energy Physics and Cosmology}, edited by
B. Kursunoglu and A. Perlmutter (American Institute of Physics,
USA).
\item [3] C. Wetterich, Nucl. Phys. B302 (1988) 668;
J.C. Jackson and M. Dodgson, Mon. Not. R. Astron. Soc. 297 (1998)
923; J.C. Jackson, Mon. Not. R. Astron. Soc. 296 (1998) 619; R.R.
Caldwell, R. Dave and P.J. Steinhardt, Phys. Rev. Lett. 80 (1998)
1582; L. Wang and P.J. Steinhardt, Astrophys. J. 508 (1998) 483;
R.R. Caldwell and P.J. Steinhardt, Phys. Rev. D57 (1998) 6057; G.
Huey, L. Wang, R. Dave, R.R. Caldwell and P.J. Steinhardt, Phys.
Rev. D59 (1999) 063005; P.F. Gonz\'{a}lez-D\'{\i}az, Phys. Rev. D62 (2000)
023513.
\item [4] S. Perlmutter {\it et al.}, Astrophys. J. 483 (1997)
565; S. Perlmutter {\it et al.} (The Supernova Cosmology Project),
Nature 391 (1998) 51; P.M. Garnavich {\it et al.} Astrophys. J.
Lett. 493 (1998) L53; B.P. Schmidt, Astrophys. J. 507 (1998) 46;
A.G. Riess {\it et al.} Astrophys. J. 116 (1998) 1009.
\item [5] A. Riess {\it et al.}, Astrophys. J. 560 (2001) 49.
\item [6] W. Fischler, A. Kashani-Poor, R. McNees and S. Paban,
JHEP 0107 (2001) 003; S. Hellerman, N. Kaloper and L. Susskind,
JHEP 0106 (2001) 003.
\item [7] T. Banks, Int. J. Mod. Phys. A46 (2001) 910; E. Witten,
{\it Quantum gravity in DeSitter space}, hep-th/0106109; T. Banks
and W. Fischler, {\it M-theory observables for cosmological
space-times}, hep-th/0102077 .
\item [8] Xiao-Gang He, {\it Accelerating universe and event
horizon}, astro-ph/0105005; S.M. Carroll, Phys. Rev. Lett. 81
(1998) 3067.
\item [9] Ya. B. Zeldovich and I.D. Novikov, Sov. Astron. A.J. 10
(1967) 602; S.W. Hawking, Mont. Not. R. Astron. Soc. 152 (1971)
75.
\item [10] F. Kottler, Ann. Phys. (Leipzig) 56 (1918) 410.
\item [11] D. Kramer, H. Stephani, M.A.H. MacCallum, and H. Herlt,
{Exact Solutions of the Einstein's Field Equations} (Cambridge
University Press, Cambridge, UK, 1980).
\item [12] G.W. Gibbons and S.W. Hawking, Phys. Rev. D15 (1977)
2738.
\item [13] P.F. Gonz\'{a}lez-D\'{\i}az, Phys. Lett. B522 (2001) 211; Phys.
Rev. D65, 104035 (2002).
\item [14] I.S. Gradshteyn and I.M. Ryzhik, {\it Table of
Integrals, Series, and Products} (Academic Press, Inc., Orlando,
USA, 1980).
\item [15] M. Abramowitz and I.A. Stegun, {\it Handbook of
Mathematical Functions} (Dover, New York, USA, 1964).
\item [16] P.J. Steinhardt, in {\it Critical Problems in Physics},
edited by V.L. Fitch and D.R. Marlow (Princeton University Press,
Princeton, N.J., USA, 1997).
\item [17] D.N. Page and S.W. Hawking, Astrophys. J. 206 (1976) 1.
\item [16] D.B. Cline, C. Motthey and S. Otwinowski, Astrophys. J.
527 (1999) 827.
\item [19] P.F. Gonz\'{a}lez-D\'{\i}az, Phys. Rev. D54 (1996) 6122; L.J.
Garay and P.F. Gonz\'{a}lez-D\'{\i}az, Gen. Rel. Grav. 33 (2001) 353.
\end{description}

\end{document}